# Mechanisms of Nanoscroll Formation and Particle Encapsulation in Janus MXenes

Sasan Rezaee[a,**], Fatemeh Mohammad Dezashibi[b], Ould el Moctar[c], Hossein Darban[d,*]

[a] Institute of Sustainable and Autonomous Maritime Systems, University of Duisburg-Essen, Duisburg, 47057, Germany
[b] Department of Physics and Energy Engineering, Amirkabir University of Technology, Tehran 159163-4311, Iran
[c] Autonomous and Energy-efficient Maritime Technologies, Technische Universität Berlin, Müller-Breslau-Str. 15, 10623 Berlin, Germany
[d] Institute of Fundamental Technological Research, Polish Academy of Sciences, Pawińskiego 5B, 02-106 Warsaw, Poland

## Abstract

Morphology transfer of 2D Janus MXenes into nanoscrolls unlocks unusual properties. Although a scalable synthesis route has been experimentally verified, the atomistic mechanism underlying nanoscroll formation remains poorly understood. We use large-scale reactive molecular dynamics simulations, validated against density functional theory (DFT) and experimental structural and elastic properties, to investigate stability and quantify the driving forces and geometry governing nanoscroll formation in three Janus MXenes, $(T_x)Ti_2C(T_y)$, where $(T_x)$ and $(T_y)$ denote the bottom and top surface terminations among bare (-b), -O, and -OH. Both square and infinitely wide flakes with lengths ranging from 10 to over 120 nm are simulated. We find that 1-7% lattice-induced strain generates a bending moment in these structures. The sheet scrolls, curves, or forms a nanotube depending on the resulting curvature and initial sheet size. For MXenes with an initial length of 120 nm, multiwalled nanoscrolls form with interlayer distances of around 0.7 nm and inner diameters of about 7 nm for $(O)Ti_2C(OH)$ and $(b)Ti_2C(OH)$, whereas $(b)Ti_2C(O)$ instead produces a much larger interlayer distance of around 1.7 nm and an inner diameter exceeding 20 nm. We show that spontaneous scrolling of a Janus MXene in the presence of an anchored nanoparticle produces a core@shell composite, in which the particle locally deforms the nanoscroll and widens the interlayer channels. This locally tunable, enlarged interlayer spacing offers a promising design route for MXene-based energy-storage electrodes. However, our simulations reveal $H_2$ gas release during encapsulation, which promotes nanobubble formation that can reduce battery life.

**Keywords:** One-dimensional MXene, Scroll, Self-rolling, Morphology, Core@shell composite.

*Corresponding authors:* *Hossein Darban; *E-mail address:* hdarban@ippt.pan.pl; *Tel.: (+48) 22 826 12 81*
**Sasan Rezaee; sasan.rezaee@uni-due.de (sasanerezaee@gmail.com)

## 1. Introduction

MXenes are a family of two-dimensional (2D) transition metal carbides, nitrides, and carbonitrides produced by selective etching of MAX phase precursors. Since their discovery in 2011 [1], MXenes have found widespread use in energy storage and harvesting, sensing, filtration, biomedicine, electromagnetic interference shielding, and antennas, owing to their exceptional combination of electronic, optical, and mechanical properties [2–5]. Despite more than fifteen years of intense research, however, most of this progress has centered on the flat, planar form of MXenes, leaving their potential in alternative morphologies largely unexplored. Since the earliest days of the field, scientists have sought a scalable, high-yield route to fabricate one-dimensional (1D) counterparts to 2D MXene flakes, inspired by rolled graphene and transition metal dichalcogenides (TMDs) [6–8]. These 1D structures, with their high aspect ratios and unique confinement effects, have enabled progress in applications ranging from tunable electronics to ion-sieving membranes [9,10].

The potential for MXenes to form 1D nanoscrolls was evident from the earliest reports, when $Ti_3C_2T_x$ nanoscrolls with radii below 20 nm were observed during MAX-phase etching [1]. Although this observation confirmed the intrinsic flexibility of MXenes for nanoscroll formation, the yield was low and uncontrolled, with scrolls forming alongside most planar flakes. Alternative nanoscroll-formation protocols, such as surfactant-assisted assembly [11,12] and freeze-drying [13], have likewise produced low quantities and inconsistent quality (i.e., without well-defined, single-scroll, channel-like structures), hindering systematic characterization and exploration of potential benefits of MXene nanoscrolls. Given theoretical predictions [14,15], the absence of a scalable, high-yield route to synthesize MXene nanoscrolls from planar flakes remained an important gap in the field.

Tubular hollow $Ti_3C_2T_x$ MXene [16] and multiwalled nanoscrolls of different materials [7,8,17,18] have been produced using sacrificial templates and spontaneous curling driven by intrinsic strain, respectively. The sacrificial-template approach requires an external sacrificial scaffold and typically produces micron-scale tubes, whereas the strain-driven approach enables the spontaneous formation of nanoscale scrolls [19,20]. Applying the intrinsic strain method yields MXene nanoscrolls by creating an asymmetric, Janus-like surface termination [21]. In this method, asymmetric surface deprotonation across the sheet establishes the lattice mismatch that drives curling. Building on this breakthrough, subsequent studies have expanded the functional scope of MXene nanoscrolls. For instance, it was demonstrated that asymmetric rolling generates an electron concentration gradient across the scroll axis that facilitates ion transport [22] and polymer composites reinforced with MXene nanoscrolls achieve synergistic enhancement in strength and toughness relative to composites reinforced with planar MXene [23].

Yet this recent body of work has so far established which conditions favor scroll formation, and what properties result, on a purely empirical basis without probing the atomistic origin of the underlying strain or the factors that dictate the resulting scroll dimensions. This scientific gap is of practical importance. Precise control over the geometry of nanotubes and nanoscrolls has proven essential for optimizing performance in high-power capacitors, battery electrodes, water-purification membranes, field-emission electron sources, and drug-delivery systems [24–29]. This gap is compounded by a second, closely related question: MXene nanoscrolls are also promising building blocks for hybrid core@shell architectures for energy-storage applications, yet realizing such composites requires understanding how a guest nanoparticle becomes encapsulated during scrolling and how that process governs the resulting interlayer structure. Atomistic simulations of scrolling and encapsulation mechanisms have already been established for other 2D materials, such as graphene and TMDs [30–35], but the corresponding mechanism for MXenes has yet to be explored.

Here, we address both gaps using large-scale reactive molecular dynamics simulations, which are validated against DFT calculations and experimental structural and elastic properties. We quantify the lattice-induced strains and the resulting bending moment in three Janus MXene structures, and establish how surface chemistry and precursor sheet geometry govern the resulting nanoscroll morphology, including its interlayer spacing and inner diameter. We further show that self-scrolling in the presence of an anchored Ti nanoparticle produces a core@shell composite with tunable, locally controlled interlayer spacing, and identify a competing $H_2$ release mechanism that must be accounted for in the design of such composites for battery applications.

## 2. Molecular dynamics simulation

### 2.1. Configuration modeling

To create Janus MXenes, we start with the conventional $Ti_2C(OH)_2$ MXene structures as shown in Figure 1 [36]. The simulations include four main classes: I: Free–Plane Strain–Free (FPSF), II: Clamped–Plane Strain–Free (CPSF), III: Free-Standing Square (FSS), and VI: CPSF supporting a single Ti nanoparticle (CPSF:TiNP@b$Ti_2C(OH)$).

In Classes I and II (see Figure 1.a), the length of $Ti_2C(OH)_2$ along the zigzag direction (x:[100]) was set to 123.15 nm, while its width along the armchair direction (y:[010]) was considered to be 15.45 nm. For these two classes, simulations were performed with periodic boundary conditions applied only along the *y*-axis, while free boundary conditions were imposed in the other directions. This allows modeling the self-scrolling of a wide flake along the zigzag direction under plane strain conditions. In Class I, both ends of the MXene sheet are free to deform, whereas in Class II, one end is fixed (i.e., atoms within 1 nm from the end are fixed in their initial positions) while the other remains free to deform.

In Class III (see Figure 1.b), we model square $Ti_2C(OH)_2$ sheets with widths of 10, 20, 40, and 100 nm to investigate the effects of initial size on the scrolling mechanism. Free boundary conditions were applied in all directions to simulate a free-standing square sheet. In Class IV (see Figure 1.c), the length and width of $Ti_2C(OH)_2$ were set to 123.15 nm and 7.70 nm along the x and y directions, respectively. Periodic boundary conditions were applied along the y direction, whereas free boundary conditions were imposed along the other directions. The MXene was clamped at one end, similar to Class II.

After preparing the initial geometry of $Ti_2C(OH)_2$ in each Class, the surface chemistry was further modified through etching to produce three distinct surface terminations: -b (i.e., bare, no termination), -OH, and -O. As a result, three different Janus MXenes, namely (O)$Ti_2C(OH)$, b$Ti_2C(OH)$, and b$Ti_2C(O)$, were created, as illustrated in Figure 1. The -b surface termination is obtained by removing both O and H atoms from the target side of the MXene, whereas the -O termination is obtained by removing only the H atoms. For the -OH termination, no modification is applied to the initial structure.

In Class IV, a spherical Ti nanoparticle (TiNP) with a diameter of 1.84 nm was created and positioned at the midpoint of the $bTi_2C(OH)$ sheet on the side without termination to construct the CPSF: TiNP@$bTi_2C(OH)$, as illustrated in Figure 1.c. The initial separation distance between the TiNP and the $bTi_2C(OH)$ surface was set equal to the van der Waals equilibrium distance between Ti atoms.

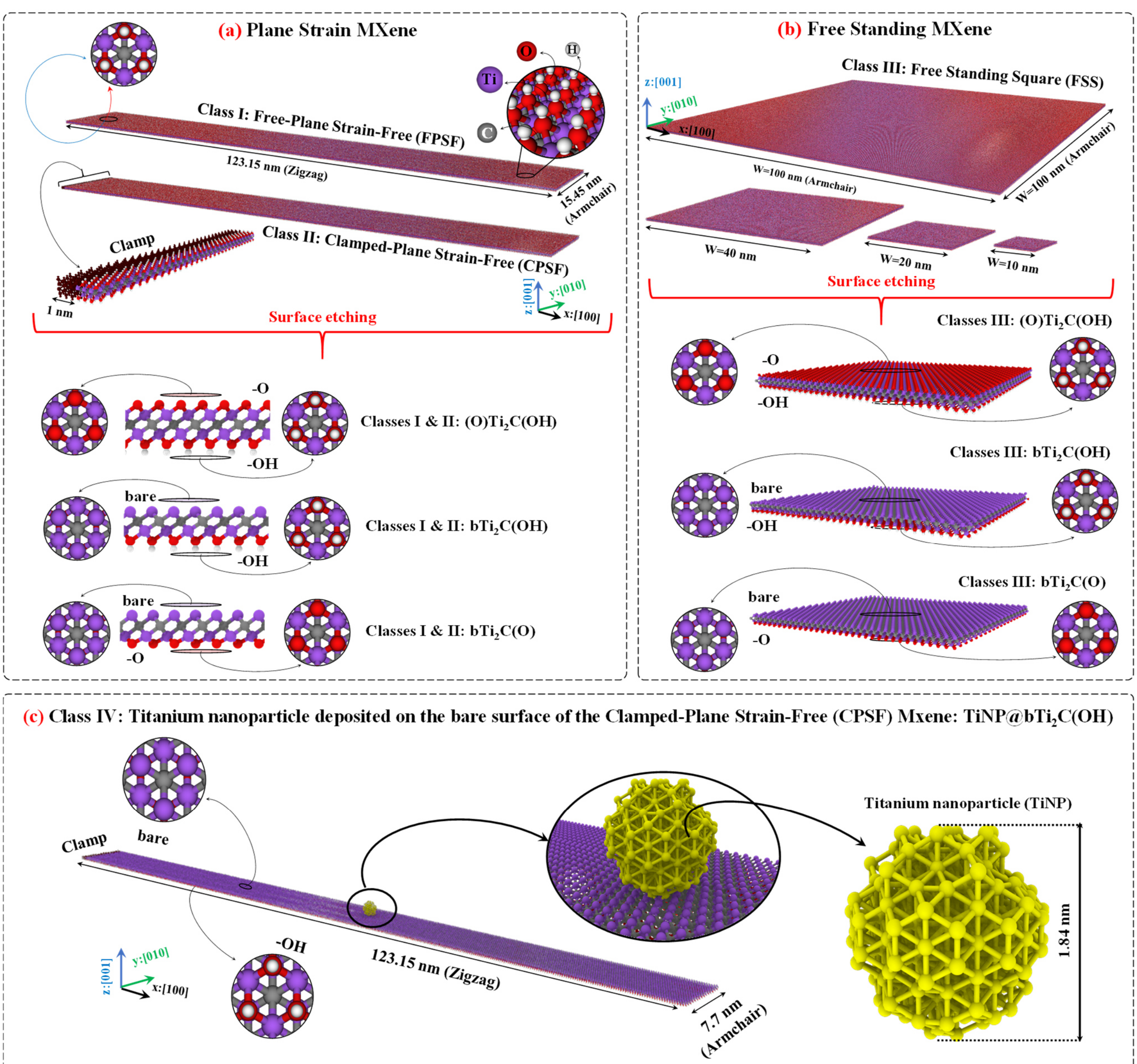

**Figure 1 The initial configurations of (a) plane strain, (b) free-standing MXene, and (c) plane strain supporting a single Ti nanoparticle. Plane strain configurations were modeled in two main classes: I: Free–Plane Strain–Free (FPSF), which means that both terminations of the MXene are free and can move freely, and II: Clamped–Plane Strain–Free (CPSF), which means that one termination is fixed as a clamp while the other termination is free. The free-standing MXene configuration consists of only one class, called III: Free-Standing Square (FSS), meaning that the four sides of the plate in the x–y plane are free to move. Three surface morphologies were considered in classes I, II, and III: $bTi_2C(O)$, $bTi_2C(OH)$, and $(O)Ti_2C(OH)$. The Clamped–Plane Strain–Free (CPSF) model supporting a single Ti nanoparticle represents Class IV. In this configuration, a single Ti nanoparticle is positioned at the midpoint of the CPSF model. The MXene surface morphology considered for Class IV was $bTi_2C(OH)$. The x:[100] and y:[010] directions are considered along the armchair and zigzag morphologies of the MXene, respectively.**

### 2.2. Simulation setups

In this work, we used the large-scale atomic/molecular massively parallel simulator (LAMMPS) software package [37] with the reactive force field (ReaxFF) [38] to study the behavior of functionalized MXene systems. The ReaxFF allows for modeling the physicochemical interactions in the system, including formation/breaking of chemical bonds. All post-processing and visualization were performed using the Open Visualization Tool (OVITO) [39]. In all simulation Classes, the time step was set to 0.5 fs. Before the dynamic simulations, energy minimization was performed using the conjugate gradient (CG) algorithm until the forces and energy converged using the tolerances of $10^{-10}$ kcal.mol$^{-1}$.Å$^{-1}$ and $10^{-10}$ (unitless), respectively. Subsequently, the canonical ensemble, commonly known as the NVT ensemble or Nosé–Hoover thermostat, was applied to the system at 1 K. During equilibration, the Janus MXenes rolled up due to energy minimization. This very low temperature helps to remove the effects of thermal fluctuations and noise in the system, thereby providing suitable conditions for the visualization of scrolling under changes in potential energy. The simulations were run for at least 75 ns to observe configuration changes and scrolling behavior. During the simulations, the mean squared displacement (MSD), radial distribution function (RDF), and potential energy per atom (PA) were calculated and monitored to reveal the mechanisms and criteria governing the rolling behavior and its characteristics. More information regarding these post-processing methods can be found elsewhere [40,41].

## 3. Results and discussions

### 3.1. Validation

To provide a benchmark for the simulation results and evaluate the ability of ReaxFF [38], as well as the simulation algorithm, in predicting the structural and mechanical attributes of MXene, the in-plane lattice constant, the thickness of the layer, the Ti–C bond length, and the directional Young's moduli of the $Ti_2C$ MXene were computed and compared with data reported in first-principles density functional theory (DFT) studies [42,43], as presented in Table 1. As this table shows, the MD simulation results accurately predict the in-plane lattice constant, the layer thickness, and the mixed covalent–metallic Ti–C bond, with errors within 1%, indicating good agreement. The potential is also able to adequately predict the Young's moduli of the MXene in the zigzag and armchair directions. This level of accuracy in predicting the structural and mechanical properties indicates that the ReaxFF potential and simulation setup can be successfully used in the rolling stage to predict the characteristics of MXene nanoscrolls, as described in the following sections. For more detailed validation of the ReaxFF potential [38] in predicting structural and mechanical properties of titanium-based MXenes, see [44].

**Table 1. Comparison of the structural and mechanical properties of $Ti_2C$ obtained in the current MD study with data reported in first-principles density functional theory (DFT) studies [42,43].**

| | Method | In-plane lattice parameter (nm) | Thickness (nm) | Ti-C bond length (nm) | Young's modulus (GPa), zigzag / armchair |
|---|---|---|---|---|---|
| Current work | MD | 0.306 | 0.229 | 0.210 | 677 / 677 |
| Ref. [42,43] | DFT | 0.303 | 0.230 | 0.209 | 620 / 600 |
| Error | - | 0.99% | 0.43% | 0.48% | 9.19% / 12.83% |

### 3.2. Plane strain

Figure 2 presents the morphological evolution, potential energy, and MSD of MXene in Class I: FPSF with the surface terminations $bTi_2C(O)$, $bTi_2C(OH)$, and $(O)Ti_2C(OH)$. As shown in the figure, all three systems undergo curvature, whereas the $bTi_2C(OH)$ and $(O)Ti_2C(OH)$ configurations form stable nanoscrolls. The observed curvature originates from two primary factors: (i) the variation in bond lengths caused by asymmetric surface terminations and (ii) the resulting bending moments induced by the mismatch in surface strain. Specifically, differences in bond lengths between the two surfaces of the MXene generate a strain mismatch by a change in the Ti-Ti sublattice, which in turn produces a net bending moment that drives the self-scrolling process.

Table 2 compares the lattice-induced strains on the surfaces of MXene with $bTi_2C(O)$, $bTi_2C(OH)$, and (O) $Ti_2C(OH)$ surface terminations after energy minimization near the end regions of the sheet. As shown, higher compressive strains are generated on the surface with the -O group in $bTi_2C(O)$, causing the Janus MXene sheet to bend toward the -O surface. Similarly, for the $bTi_2C(OH)$ and $(O)Ti_2C(OH)$ systems, the compressive strains on the -b and -O surfaces, respectively, relative to the tensile strain on the -OH surface, induce bending of the nanosheets toward the -b and -O terminations. The underlying mechanism of this curvature, governed by the induced net bending moment, is discussed in detail in Section 3.5.

**Table 2 Comparison of the lattice-induced strains in surfaces of MXene with different surface terminations: $bTi_2C(O)$, $bTi_2C(OH)$, and $(O)Ti_2C(OH)$, resulting from surface etching near the free tail and head of the sheet.**

| Surface termination | Ti-Ti (nm) | | Strain[1] | | Type of strain | | Net strain | Bend toward |
|---|---|---|---|---|---|---|---|---|
| $bTi_2C(O)$ | -b | -O | -b | -O | -b | -O | 4.5% | -O |
| | 0.299 | 0.285 | -2.6% | -7.1% | Compressive[2] | Compressive | | |
| $bTi_2C(OH)$ | -b | -OH | -b | -OH | -b | -OH | 7.4% | -b |
| | 0.297 | 0.320 | -3.2% | 4.2% | Compressive | Tensile | | |
| $(O)Ti_2C(OH)$ | -O | -OH | -O | -OH | -O | -OH | 5.2% | -O |
| | 0.302 | 0.318 | -1.6% | 3.6% | Compressive | Tensile | | |

[1]The initial Ti–Ti bond length was 0.307 nm. The strain was then calculated as [(final bond length-initial bond length)/(initial bond length)].

[2]Both the -b and -O surface terminations experience compressive strain; however, the higher compressive strain in the -O surface termination compensates for the compressive strain on the -b surface, resulting in a net strain of 4.5% in the system.

Figure 2 demonstrates that the curvature and scrolling morphologies of MXene are strongly dependent on the surface termination. After energy minimization, the I:FPSF: $bTi_2C(O)$ configuration experiences a strain gradient across the thickness with a net value of 4.5%, which causes the scrolling process. However, due to the specific surface terminations and the limited variation in Ti-C bond lengths on both the -b and -O surfaces, the I:FPSF:$bTi_2C(O)$ system tends to form a large curvature with a relatively large diameter. Consequently, two large rings are formed in the MXene nanosheet, with a diameter of 22.15 nm and an interlayer distance of 1.75 nm (Table 3). This configuration is metastable because, as the simulation progresses, interlayer zipping occurs due to Coulombic interactions between the Ti in the scroll's inner layer and the O in the outer layer (for further insights, see the simulation results and discussions for Class II). In contrast, for I:FPSF:b$Ti_2C(OH)$ and I:FPSF:(O)$Ti_2C(OH)$, stable nanoscrolling is achieved, and the resulting configurations remain stable throughout the simulation. Table 3 compares the diameter of the innermost ring and the interlayer distance of MXene in Class I, showing that the innermost ring diameter follows the order: b$Ti_2C(O)$ > (O)$Ti_2C(OH)$ > b$Ti_2C(OH)$. This observation is in agreement with the net strains reported in Table 2 for these Janus MXenes.

**Table 3 Comparison of the interlayer distance and the diameter of the innermost ring of MXene nanoscrolls in class I: FPSF with different surface terminations: b$Ti_2C(O)$, b$Ti_2C(OH)$, and (O)$Ti_2C(OH)$.**

| Class | Surface termination | Interlayer distance (nm) | Diameter of innermost ring (nm) | Stability status |
|---|---|---|---|---|
| I: FPSF | b$Ti_2C(O)$ | 1.75 | 22.15 | Metastable |
| I: FPSF | b$Ti_2C(OH)$ | 0.65 | 6.00 | Stable |
| I: FPSF | (O)$Ti_2C(OH)$ | 0.70 | 7.10 | Stable |

Figure 2.d and e compare the evolution of the potential energy and MSD of Class I systems, respectively, over a 1 ns simulation. Figure 2.d clearly shows that the system's potential energy decreases as scrolling initiates, indicating the intrinsic tendency of the Janus MXene sheets to self-scroll due to energy minimization. Figure 2.e illustrates the MSD of Class I along the z direction ($MSD_z$), as the dimension of the system along this axis changes during the scrolling process. As shown, the $MSD_z$ increases after the initiation of scrolling. However, after a specific time, which depends on the surface termination, a noticeable change in the $MSD_z$ trend is observed, represented by a smooth decrease in the diagram. By correlating this point with the snapshots in Figure 2.a–c, this decrease corresponds to the stage at which the MXene sheet reaches its first curved ring. After this stage, the formation of the nanoscrrolled configuration can be confirmed. The MSD profiles further demonstrate that, among the Class I MXenes with different surface terminations, (O)$Ti_2C(OH)$ reaches the first scrolled configuration earlier than b$Ti_2C(O)$ and b$Ti_2C(OH)$.

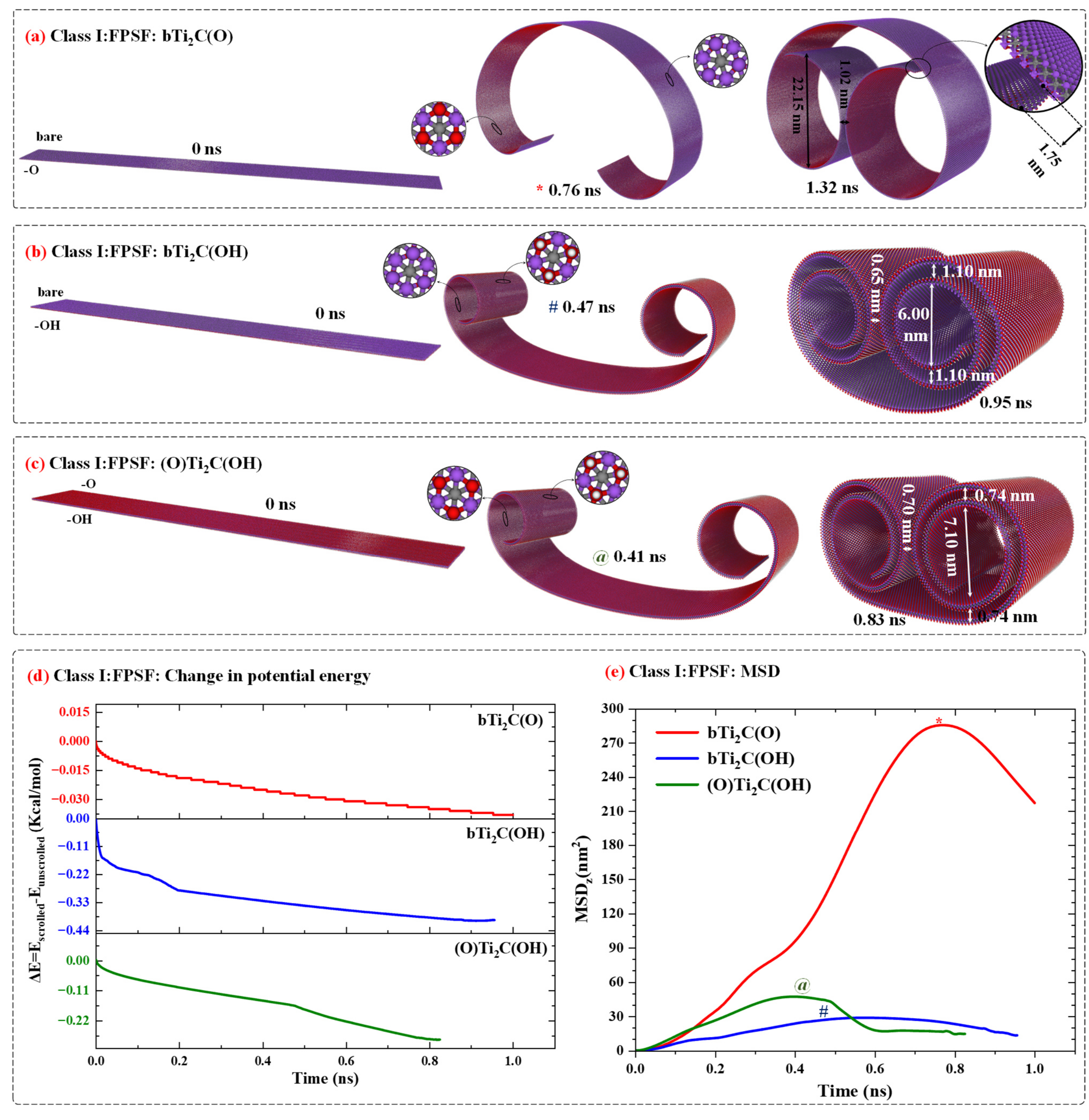

**Figure 2** Scrolling snapshots of Class I: FPSF with different surface terminations: **(a)** $bTi_2C(O)$, **(b)** $bTi_2C(OH)$, and **(c)** $(O)Ti_2C(OH)$. **(d)** Evolution of the potential energy and **(e)** mean squared displacement (MSD) along the z:[001] direction during the 1 ns simulation.

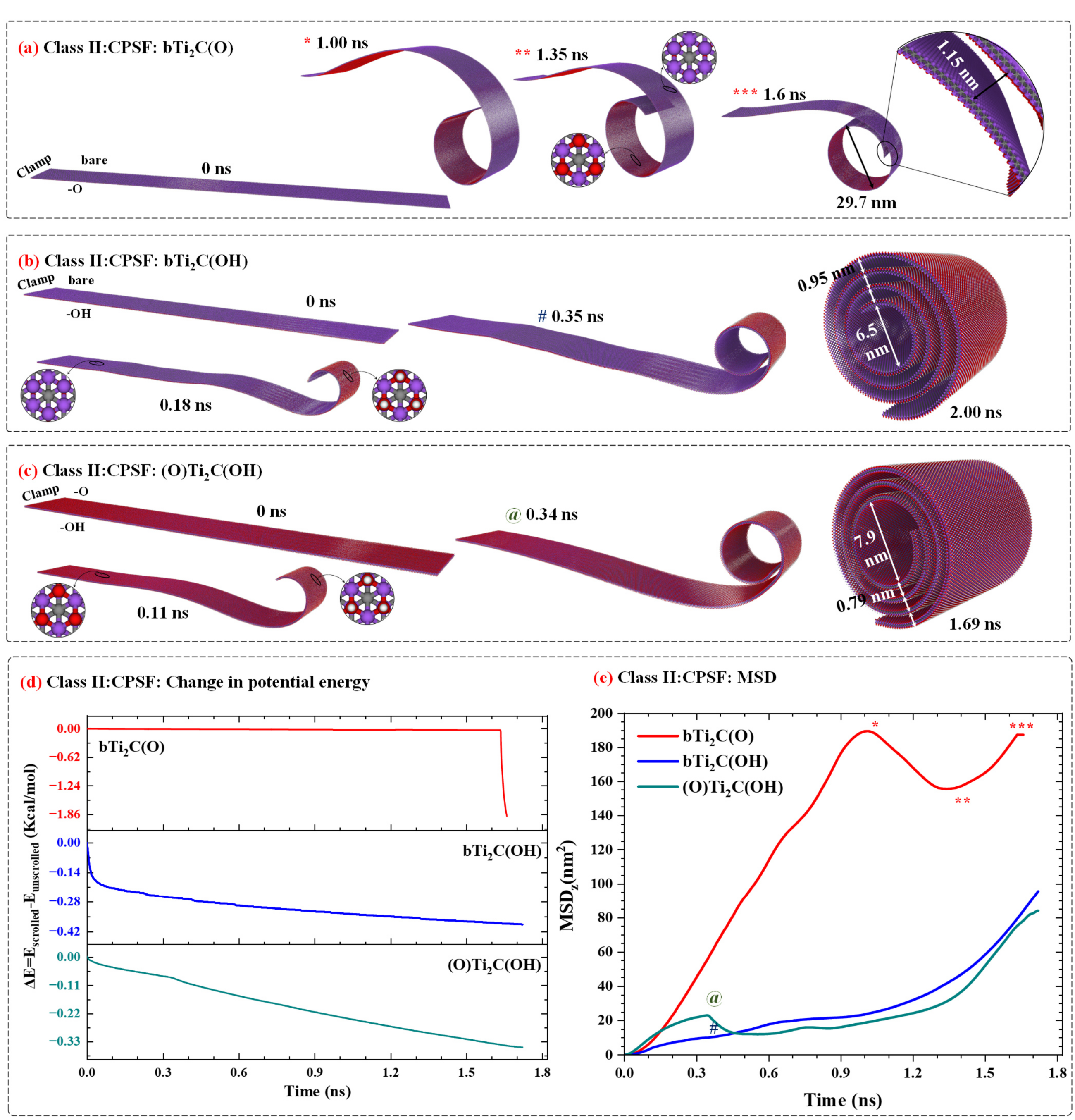

**Figure 3** **Scrolling snapshots of Class II: CPSF with different surface terminations: (a) $bTi_2C(O)$, (b) $bTi_2C(OH)$, and (c) $(O)Ti_2C(OH)$. (d) Evolution of the potential energy and (e) mean squared displacement (MSD) along the z:[001] direction during the 1 ns simulation.**

Figure 3 demonstrates the evolution of nanoscrolling morphology, potential energy, MSD of MXene in Class II:CPSF with the surface terminations $bTi_2C(O)$, $bTi_2C(OH)$, and $(O)Ti_2C(OH)$. The scrolling behavior, potential energy variation, and MSD evolution in this class exhibit trends similar to those observed in Class I. In both Classes I and II, the $bTi_2C(O)$, $bTi_2C(OH)$, and $(O)Ti_2C(OH)$ systems roll toward the -O, -b, and -O surface terminations, respectively. Additionally, the decrease in potential energy during nanoscrolling highlights the system's intrinsic tendency to scroll. The MSD profiles also show a characteristic decrease corresponding to the formation of the first curvature ring. Similar to Class I, the (O)$Ti_2C$(OH) system in Class II reaches its first curvature ring earlier than the other two configurations. The strains induced during the scrolling of MXene sheets in Class II are consistent with the values reported in Table 2.

Table 4 summarizes the interlayer distance and the diameter of the innermost ring of MXene nanoscrolls in Class II:CPSF with the surface terminations b$Ti_2C$(O), b$Ti_2C$(OH), and (O)$Ti_2C$(OH). A comparison of these results with those reported for Class I:FPSF in Table 3 reveals that introducing a single clamp at one end of the MXene sheet results in a moderately larger innermost ring. Specifically, the innermost ring diameters of the CPSF systems with b$Ti_2C$(O), b$Ti_2C$(OH), and (O)$Ti_2C$(OH) surface terminations are approximately 34%, 8%, and 11% larger than those of the corresponding I:FPSF systems, respectively. Furthermore, the interlayer distances of the scrolled II:CPSF systems with b$Ti_2C$(OH) and (O)$Ti_2C$(OH) surface terminations are approximately 46% and 13% greater than those of the corresponding I:FPSF systems. These results indicate that clamping one end of the MXene sheet has a measurable influence on the scrolling diameter and interlayer spacing. This is primarily because the Class II configuration, owing to its clamped end, effectively represents a Janus MXene with twice the length of the MXenes modeled in Class I. This difference in the resulting nanoscroll geometry highlights its dependence on the initial length of the MXene precursor.

The b$Ti_2C$(O) systems in Classes I and II exhibit similar nanoscrolling mechanisms; however, unlike the other two surface terminations (b$Ti_2C$(OH) and (O)$Ti_2C$(OH)), they do not show a consistent trend in interlayer distance. This behavior is attributed to the metastable nature of the b$Ti_2C$(O) configuration in both classes, which prevents the formation of a well-defined nanoscroll and consequently yields less reliable values for the interlayer distance and the diameter of the innermost ring as the system transitions from Class I to Class II. The term "metastable" is used to describe the semi-nanoscrolled b$Ti_2C$(O) configurations in Classes I and II because they cannot maintain their partially scrolled morphology over time. After the formation of the first ring, interlayer zipping occurs, causing the semi-nanoscrolled structure to collapse into a more energetically favorable configuration. The occurrence of interlayer zipping can be attributed to two factors. First, the variation in Ti-C bond lengths is insufficient to generate the bending moment required to sustain a stable curvature. Second, the strong Coulombic attraction between Ti and O atoms in adjacent layers initiates interlayer zipping, resulting in a rapid decrease in the system's potential energy. This behavior is clearly illustrated by the II:CPSF:b$Ti_2C$(O) potential energy drop in Figure 3.d.

**Table 4 Comparison of the interlayer distance and the diameter of the innermost ring of MXene nanoscrolls in class II:CPSF with different surface terminations: b$Ti_2C$(O), b$Ti_2C$(OH), and (O)$Ti_2C$(OH).**

| Class | Surface termination | Interlayer distance (nm) | Diameter of innermost ring (nm) | Stability status |
|---|---|---|---|---|
| II: CPSF | b$Ti_2C$(O) | 1.15 | 29.70 | Metastable |
| II: CPSF | b$Ti_2C$(OH) | 0.95 | 6.50 | Stable |
| II: CPSF | (O)$Ti_2C$(OH) | 0.79 | 7.90 | Stable |

### 3.3. Free-standing square (FSS)

Figure 4 illustrates the snapshots of Class III:FSS during scrolling for different sheet sizes and surface terminations, including $bTi_2C(O)$, $bTi_2C(OH)$, and $(O)Ti_2C(OH)$. The figure clearly shows the same bending and scrolling mechanisms as those identified for Classes I:FPSF and II:CPSF. All FSS configurations curve toward the surface termination that experiences a more compressive strain. The FSS structures with morphologies of $bTi_2C(O)$, $bTi_2C(OH)$, and $(O)Ti_2C(OH)$ bend toward the -O, -b, and -O surface terminations, respectively. Figures 4.a and 4.b demonstrate that FSS structures primarily develop curvature, while only $bTi_2C(OH)$ can form an additional morphology similar to a chiral nanotube. For MXene sheets with widths of 10 and 20 nm, the surface termination mismatch initiates scrolling; however, the final morphology of the bent sheet is constrained by the sheet size. In the case of $bTi_2C(OH)$ with a width of 20 nm, the bending moment generates a suitable curvature diameter that is not restricted by the sheet dimensions, enabling the formation of a nanotube-like structure. Figures 4.c and 4.d show that increasing the sheet size beyond 20 nm enhances the possibility of scrolling; however, $bTi_2C(O)$ exhibits only a curved morphology and cannot form a nanoscrolled structure. In contrast, $bTi_2C(OH)$ and $(O)Ti_2C(OH)$ with sheet sizes of 40 and 100 nm can form nanoscrolled MXene structures because the sheet dimensions are sufficiently large to accommodate scrolling under the curvature induced by the bending moment.

Table 5 compares the nanoscrolling ability of Class III: FSS MXene as a function of surface termination. This table clearly shows that three distinct morphologies: curvature, nanotube, and nanoscroll, can be obtained in FSS MXene, and their formation is governed by both the sheet size and the surface termination. All configurations exhibit bending and develop curvature because it tends to reduce potential energy. However, as the sheet length (and width) increases, the possibility of nanotube and nanoscroll formation also increases. For $bTi_2C(O)$, curvature is observed for all sheet sizes, which is associated with the metastable nature of this MXene. In the case of $bTi_2C(OH)$, the 20 nm sheet bends and, through the formation of curvature, its opposite edges merge to form a chiral MXene nanotube. Increasing the sheet size beyond 20 nm further promotes scrolling, resulting in nanoscroll formation, as observed for the 40 and 100 nm sheets. Therefore, these results demonstrate that MXene morphology can be tailored to form curved sheets, nanotubes, or nanoscrolls by controlling sheet size and surface termination.

**Table 5 Comparison of the nanoscrolling ability of Class III: FSS MXene as a function of surface termination and MXene size.**

| Class | Surface termination | Nanoscrolling ability with respect to the size of MXene | | | |
|---|---|---|---|---|---|
| | | 10 (nm) | 20 (nm) | 40 (nm) | 100 (nm) |
| FSS: III | $bTi_2C(O)$ | Curvature | Curvature | Curvature | Curvature |
| FSS: III | $bTi_2C(OH)$ | Curvature | Nanotube | Nanoscroll | Nanoscroll |
| FSS: III | $(O)Ti_2C(OH)$ | Curvature | Curvature | Nanoscroll | Nanoscroll |

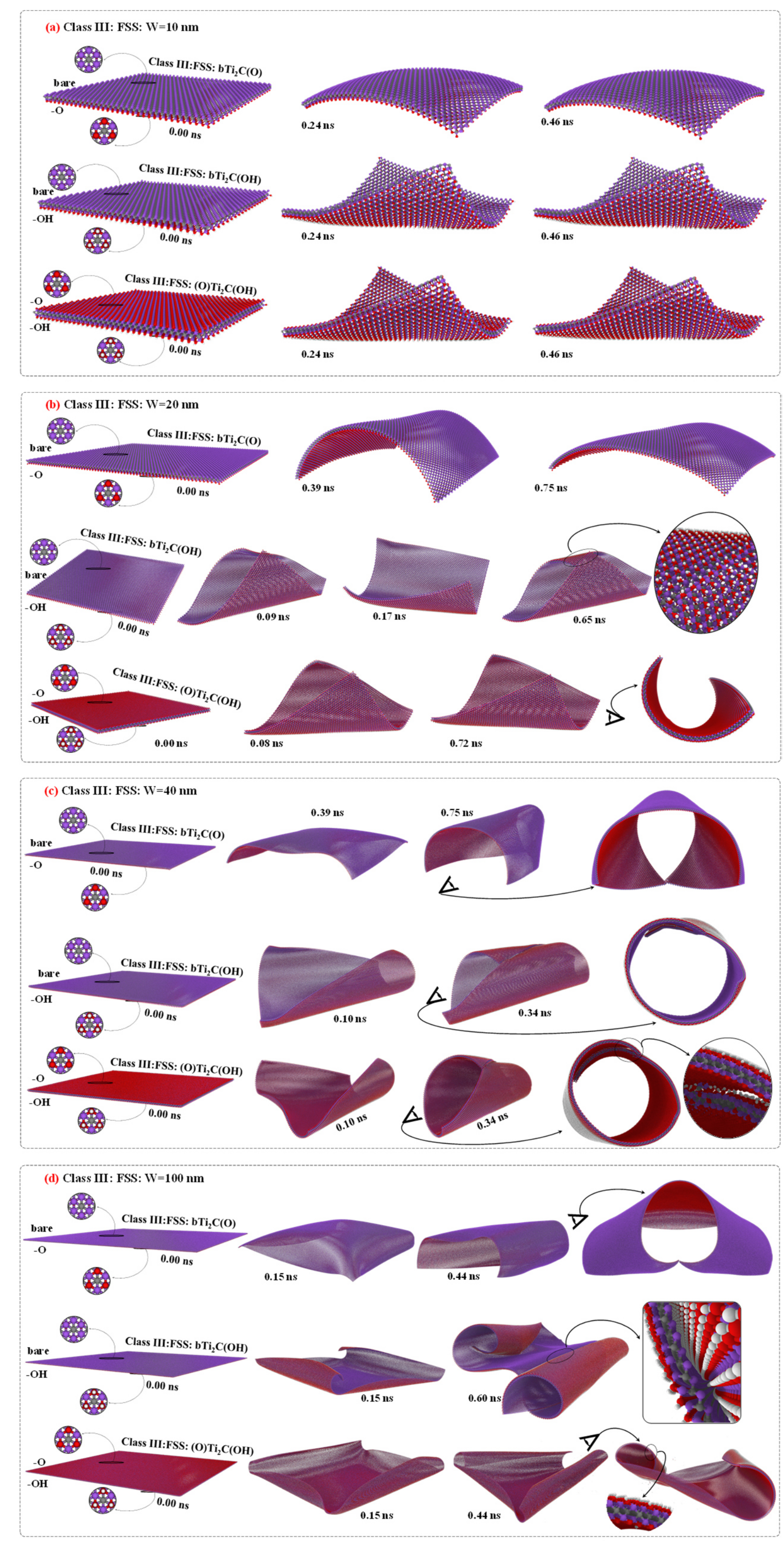

**Figure 4 Scrolling snapshots of Class III:FSS with different surface terminations of $bTi_2C(O)$, $bTi_2C(OH)$, and $(O)Ti_2C(OH)$ as function of sheet size. The configurations have a width-to-length ratio of 1. Four configurations with (w=width=length) were considered: (a) w = 10 nm, (b) w = 20 nm, (c) w = 40 nm, and (d) w = 100 nm.**

Figure 5 compares the changes in the potential energy and the MSD along the z direction for Class III: FSS MXene with different surface terminations, including $bTi_2C(O)$, $bTi_2C(OH)$, and $(O)Ti_2C(OH)$, as a function of sheet size. The trends observed in both the potential energy and MSD curves are consistent with those obtained for the previous MXene classes (I and II). As the simulation progresses and the MXene sheets bend, the potential energy decreases, indicating the intrinsic tendency of the FSS MXene sheets to curve. Similarly, the MSD increases with the onset of bending, reflecting the morphological evolution of the sheets along the z direction. However, for the FSS MXene sheets with sizes of 10, 20, and 40 nm, the MSD curves become nearly horizontal with small oscillations after approximately 0.25 ns. This behavior indicates that the sheet morphology has reached a stable configuration and undergoes no further significant changes with time. In contrast, for the 100 nm sheet, additional morphological evolution is expected if the simulation continued. A similar trend is observed in the potential energy curves. For the 10, 20, and 40 nm sheets, the potential energy also reaches a nearly horizontal oscillatory profile after approximately 0.25 ns, indicating that the systems have attained equilibrium and that no further significant morphological changes are expected.

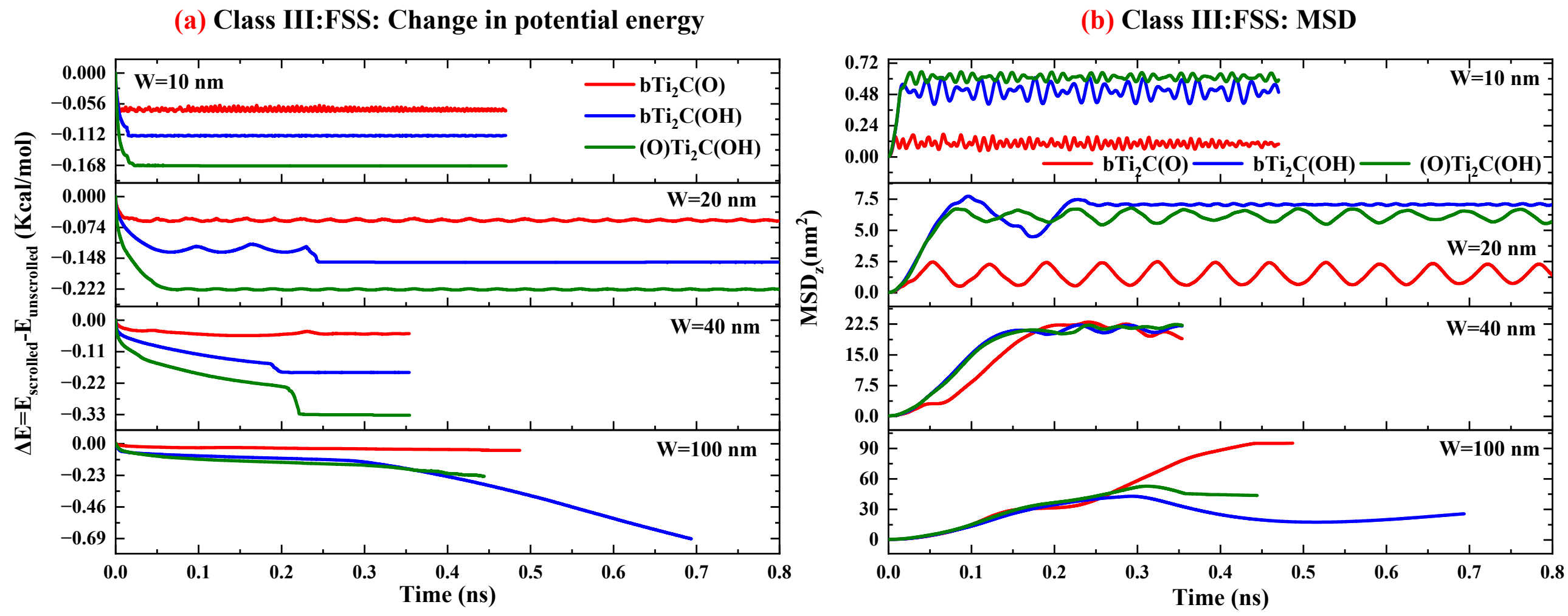

**Figure 5 Change in potential energy and mean squared displacement (MSD) of class III:FSS with different surface terminations of $bTi_2C(O)$, $bTi_2C(OH)$, and $(O)Ti_2C(OH)$ as a function of sheet size. (a) Evolution of the potential energy and (b) MSD along the z:[001] direction during the 1 ns simulation.**

### 3.4. Self-rolling mechanism

To elucidate the mechanism of self-rolling, snapshots of II:CPSF:$(O)Ti_2C(OH)$ over the initial 0.710 ns are presented in Figure 6. In this figure, the Ti-O bond lengths, O~O, and Ti~Ti distances are shown to illustrate the underlying mechanisms responsible for self-rolling. This figure clearly shows that, in the flat region, the Ti-O bond length is 1.95 Å for the surface terminated with -OH groups, whereas it decreases to 1.92 Å for the -O terminated surface after the H atoms are etched. These bond lengths were calculated from the RDF diagrams and represent the bond lengths in regions away from the curved areas. A comparison of the Ti-O bond lengths between the -OH and -O terminated surfaces (Figure 6 after minimization energy) indicates that the removal of H atoms reduces the Ti-O bond length by approximately 1.54%, thereby inducing linear expansion in the -OH surface layer while causing linear contraction in the -O terminated surface. In other words, tensile and bending stresses are generated in the -OH and -O surface layers, respectively, following the etching of H atoms from one surface. Consequently, the driving forces required for rolling already exist within the layers, and only an initiating force is needed to trigger rolling toward the surface exhibiting the shorter Ti-O bond. The free head at the edge of the II:CPSF:(O)$Ti_2C$(OH) sheet provides this initiating force, allowing the rolling process to begin.

This simulation setup resembles the actual synthesis route developed in [21]. The method relies on controlled aqueous delamination, using cation intercalation (e.g., $Li^+$ or tetramethylammonium) combined with mechanical agitation under conditions tuned to favor scroll formation. Starting from few-layer MXene stacks, the process proceeds iteratively, with individual layers peeling off one by one and rolling into a scroll. In this method, asymmetric surface deprotonation across the sheet establishes the lattice mismatch that drives rolling.

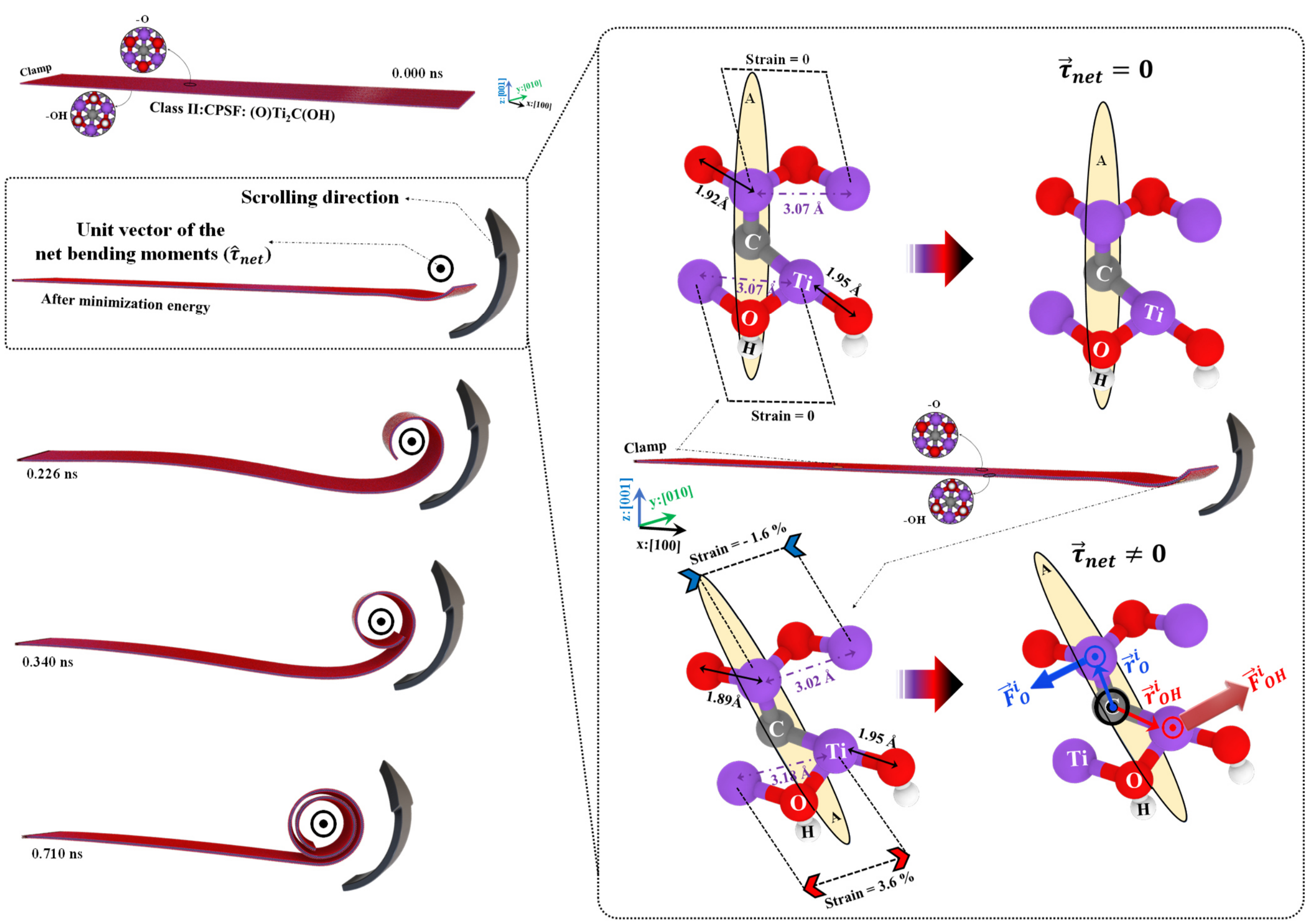

**Figure 6 Snapshots of the II:CPSF:(O)$Ti_2$C(OH) sheet during self-rolling over a period of 0.710 ns. In this figure, the unit vector of the net bending moment, along with the scrolling direction, is illustrated. A magnified view of the system following energy minimization is shown on the right-hand side, where the origin of the scrolling process is identified within the C–Ti–O–Ti–O and C–Ti–OH–Ti–OH atomic chains, with the C atoms serving as the pivot points.**

The strain-induced bending can be quantified using both full MD and analytical–MD methods. In the full MD approach, the strain, forces, and bending moment are directly calculated from the MD trajectories. In the analytical–MD approach, the strain and position vectors are calculated from the MD simulations and then integrated into physics-based equations to determine the net bending moment responsible for curvature and scrolling. In this study, both the MD and analytical–MD methods were employed and yielded comparable results with the same order of magnitude. The analytical–MD method is described in the following section to provide more physical insights.

The difference in Ti-O bond lengths on either side of the Janus MXene leads to changes in the sublattice arrangement of O and Ti atoms. As a result, the O~O and Ti~Ti interatomic distances in the outer and inner layers of the scroll are altered, inducing a torque, specifically referred to here as a bending moment. To provide a clear visualization of the net bending moment in the system, Figure 6 presents magnified views of the C~Ti~O~Ti~O and C~Ti–OH–Ti–OH atomic chains, with the C atoms serving as pivot points. This figure clearly demonstrates that the variation in Ti-O bond lengths on the -O and -OH termination surfaces generates net strains of -1.6% and 3.6% in these respective layers. These conditions can be considered equivalent to a net tensile strain of 5.2% in the -OH surface termination layer. The bending moments ($\vec{\tau}_{net}$) generated by the asymmetric -O and -OH surface terminations can be expressed as follows:

$$\vec{\tau}_{net} = \sum_{i=1}^{N} \vec{\tau}_C^i \quad (1.a)$$

$$\vec{\tau}_C^i = \sum_{j=1}^{3} \vec{\tau}_{OH}^{ji} + \sum_{k=1}^{3} \vec{\tau}_O^{ki} = \vec{\tau}_{OH}^{yi} + \vec{\tau}_O^{yi} \quad (1.b)$$

Here, $N$ denotes the total number of all central carbon atoms, and $\vec{\tau}_i^C$ represents the net bending moment applied to the i[th] C atom. Because each C is bonded to six Ti atoms, three associated with the -O surface termination and three associated with the -OH surface termination, the $\vec{\tau}_C^i$ can be expressed as $\sum_{j=1}^{3} \vec{\tau}_{OH}^{ji} + \sum_{k=1}^{3} \vec{\tau}_O^{ki}$. Here, j and k denote the number of Ti atoms connected to the i[th] C atom in the -OH and -O surface termination layers, respectively. $\vec{\tau}_{OH}^{j}$ and $\vec{\tau}_O^{k}$ are bending moments in -OH and -O surface termination layers, respectively. However, since the strain acts only along the x:[100] direction, only the bending moments generated by force components in this direction contribute to the self-rolling behavior. Consequently, the equation can be simplified by considering only the bending moments induced by forces acting along the x direction in the -OH and -O surface terminations, denoted as $\vec{\tau}_{OH}^{y}$ and $\vec{\tau}_O^{y}$, respectively. The superscript y denotes the direction of the bending moment about the y:[010] axis, as determined according to the right-hand rule. These bending moments can be expressed in terms of the strain-induced forces and the position vectors measured from the pivot point, as follows:

$$\vec{\tau}_C^i = \vec{\tau}_{OH}^{yi} + \vec{\tau}_O^{yi} = \left(\vec{r}_{OH}^i \times \vec{F}_{OH}^i\right) + \left(\vec{r}_O^i \times \vec{F}_O^i\right) \quad (2)$$

Here, $\vec{r}_{OH}^{\,i}$ and $\vec{r}_{O}^{\,i}$ represent the position vectors measured from the line of action of the forces in the -OH and -O surface terminations, respectively. $\vec{F}_{OH}^{i} = YA|\varepsilon_{OH}^{i}|\,\hat{F}_{OH}^{i}$ denotes the generated force in the -OH surface termination due to the lattice-induced strains (see Table 2), whereas $\vec{F}_{O}^{i} = YA|\varepsilon_{O}^{i}|\,\hat{F}_{O}^{i}$ represents the corresponding force in the -O surface termination. $\hat{F}_{OH}^{i}$ and $\hat{F}_{O}^{i}$ are the unit vectors indicating the directions of the strain-induced forces at the -OH and -O surface terminations, respectively. The strains ($\varepsilon_{OH}^{i}$ and $\varepsilon_{O}^{i}$) were expressed in absolute terms ($|\varepsilon_{OH}^{i}|$ and $|\varepsilon_{O}^{i}|$) because the corresponding unit vectors inherently account for their signs (positive or negative). Based on Figure 6, the unit vectors are parallel but oriented in opposite directions.

$Y$ represents the Young's modulus of MXene (ranging from 330 to 677 GPa) [42,45,46], $A$ denotes the nominal cross-sectional area, and $\varepsilon$ represents the lattice-induced strain in -OH ($\varepsilon_{OH}^{i} = 0.036$) and -O ($\varepsilon_{O}^{i} = -0.016$) surface termination (see Table 2). Considering that the average $\varepsilon_{OH}^{i}$ can be approximately 2.25 times greater than $\varepsilon_{O}^{i}$, it can be rewritten as $|\varepsilon_{OH}^{i}| = 2.25|\varepsilon_{O}^{i}|$. The cross-sectional area $A$ can be defined as ($\pi r^2$), where r is approximately 2.5 Å, which is chosen to be slightly larger than the Ti-C bond length (Table 1) to ensure that the entire atomic chains are included within the nominal cross-sectional area. The values of $\vec{r}_{OH}^{\,i}$ and $\vec{r}_{O}^{\,i}$ are approximately equal to the Ti-C bond length; therefore, they can be expressed as $\vec{r}_{OH}^{\,i} = d_{Ti-C}\,\hat{r}_{OH}^{i}$ and $\vec{r}_{O}^{\,i} = d_{Ti-C}\,\hat{r}_{O}^{i}$, where $d_{Ti-C} = 2.1$ Å represents the Ti-C bond length, $\hat{r}_{OH}^{i}$ and $\hat{r}_{O}^{i}$ are the unit vector of the first Ti connected to C pivot point in -OH and -O surface terminations, respectively. Based on these specified variables, Eq. (2) can be rewritten as follows:

$$\vec{\tau}_{C}^{\,i} = \tau_{(O-OH)}\left(\left(2.25\,\hat{r}_{OH}^{i} \times \hat{F}_{OH}^{i}\right) + \left(\hat{r}_{O}^{i} \times \hat{F}_{O}^{i}\right)\right) \quad (3.a)$$

$$\vec{\tau}_{C}^{\,i} = \tau_{(O-OH)}(2.25\ \sin\theta_{OH} + \sin\theta_{O})(-\hat{j}) \quad (3.b)$$

Here, $\tau_{(O-OH)} = Y\pi r^2|\varepsilon_{O}^{i}|\,d_{Ti-C}$ and is in the range of $2.18\times10^{-19}$ N.m to $4.46\times10^{-19}$ N.m. The average value of $\tau_{(O-OH)}$, denotes as $\bar{\tau}_{(O-OH)}$, can be defined as $3.32\times10^{-19}$ N.m. $\theta_{OH}$ is the angle between $\hat{r}_{OH}^{i}$ and $\hat{F}_{OH}^{i}$, $\theta_{O}$ denotes the angle between $\hat{r}_{O}^{i}$ and $\hat{F}_{O}^{i}$, and $\hat{j}$ points to the unit vector along the Cartesian y:[010] direction. Eq. (3.b) clearly reveals that the bending moments induced by the -OH and -O surface terminations are oriented in the same direction, resulting in rolling along the black arrow according to the right-hand rule, as depicted in Figure 6. For each central C atom, the MD calculations reveal that the net bending moment can be on the order of $10^{-19}$ N.m. This small value, generated due to the change in Ti-O bond lengths and the lattice mismatch strain between the -OH and -O surface terminations, acts as the initiating bending moment that drives (O)$Ti_2$C(OH) scrolling. The magnitude of the bending moment will vary with changes in temperature and surface termination; however, it remains the primary driving force responsible for the scrolling process. The origin of scrolling for other surface terminations of b$Ti_2$C(O) and b$Ti_2$C(OH) is also associated with bending moments; however, their magnitudes are different. Using the same approach, it can be shown that the net bending moments acting on the C pivot atom for the Class II MXene systems, denoted as II:CPSF:b$Ti_2$C(O) and II:CPSF:b$Ti_2$C(OH), are given as follows:

$$\vec{\tau}_C^i = \tau_{(b-O)}(-\sin\theta_b + 2.73\sin\theta_O)(\hat{j}) \quad (4)$$

$$\vec{\tau}_C^i = \tau_{(b-OH)}(\sin\theta_b + 1.31\sin\theta_{OH})(-\hat{j}) \quad (5)$$

Here, $\tau_{(b-O)} = Y\pi r^2|\varepsilon_b^i|\, d_{Ti-C}$ and $\tau_{(b-OH)} = Y\pi r^2|\acute{\varepsilon}_b^i|\, d_{Ti-C}$ represent the initial net bending moments corresponding to $bTi_2C(O)$ and $bTi_2C(OH)$, respectively. $|\varepsilon_b^i|$ is 0.026 and $|\acute{\varepsilon}_b^i|$ equal to 0.032 (see Table 2). $\theta_b$ is the angle between $\hat{r}_b^i$ and $\hat{F}_b^i$. Figure 7 compares the magnified views of the C~Ti~O~Ti~O and C~Ti~OH~Ti~OH atomic chains, along with their strain-induced forces and corresponding bending moments, for the Class II MXene systems with $bTi_2C(O)$, $bTi_2C(OH)$, and $(O)Ti_2C(OH)$ surface terminations. Substituting the specific variables into Eqs. (4) and (5) shows that the initial average bending moments corresponding to $bTi_2C(O)$ and $bTi_2C(OH)$ are in the order of 5.39 $\times 10^{-19}$ N.m and 6.64 $\times 10^{-19}$ N.m, respectively. Intercomparison of these values with $\bar{\tau}_{(O-OH)}$ shows that the average net bending moment induced by the surface terminations in Class II: CPSF follows the sequence: $bTi_2C(OH) > bTi_2C(O) > (O)Ti_2C(OH)$. As the magnitude of the net bending moment is determined by defining the nominal cross-sectional area and depends on the atomic positions of the C atoms, its absolute values may be subject to uncertainty; however, the observed trends are reliable and reproducible.

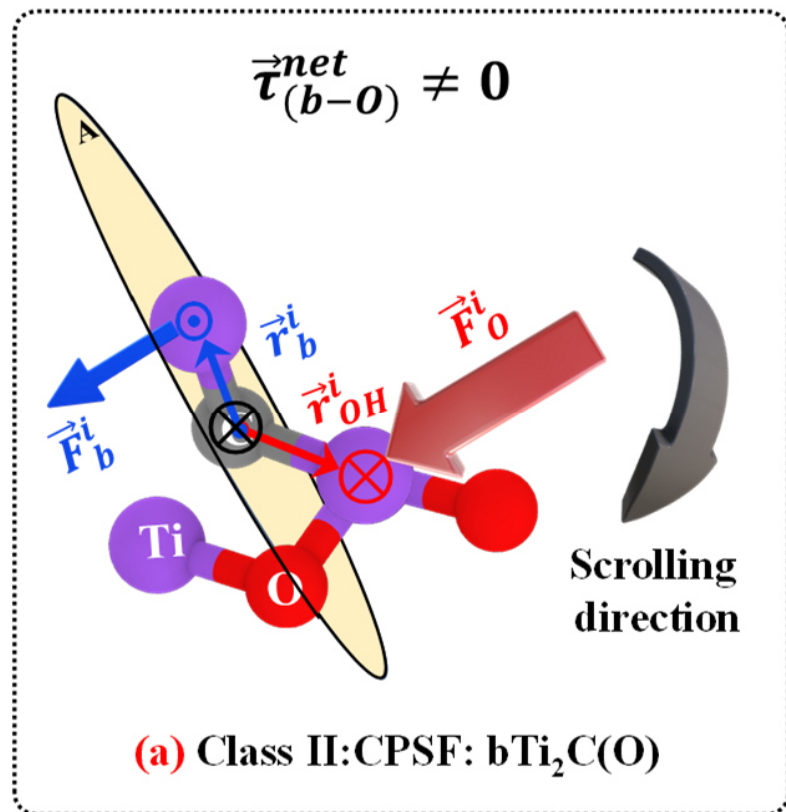

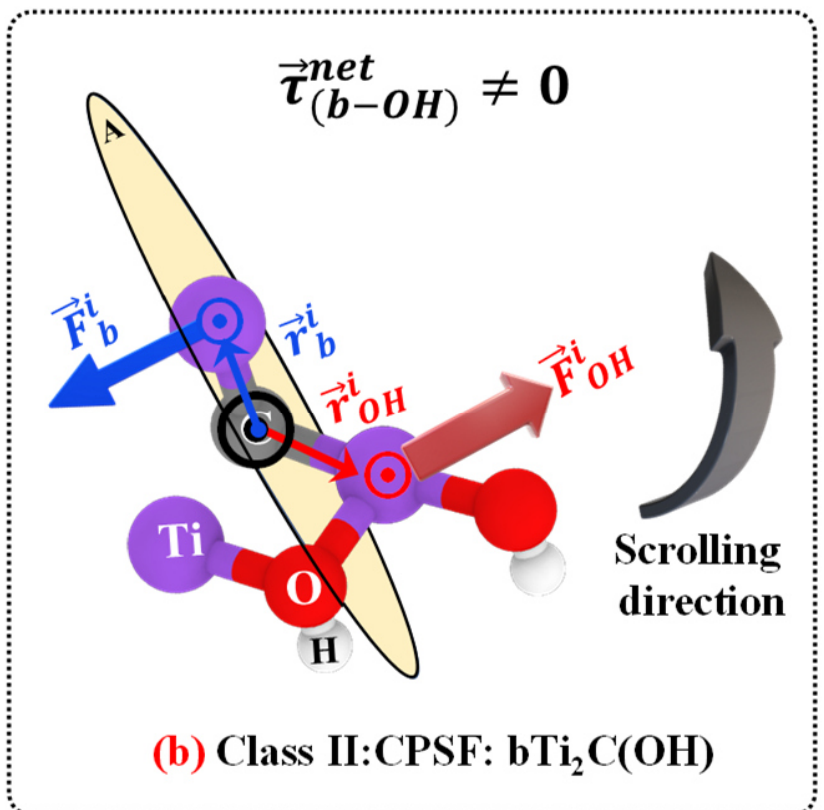

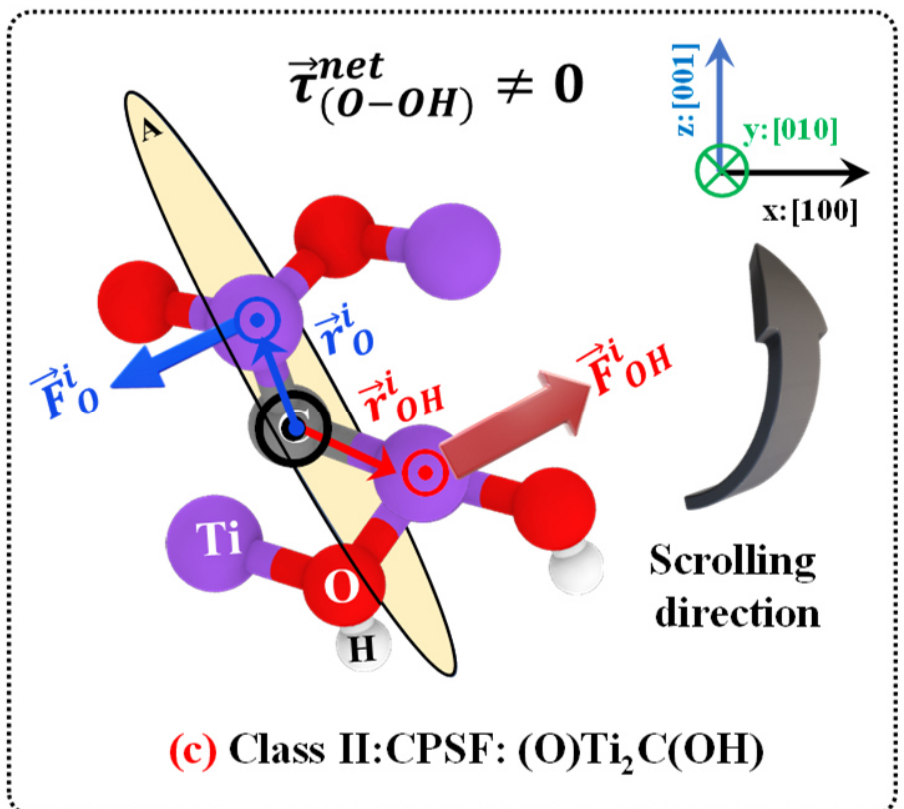

**Figure 7 Comparison of the induced strain, the corresponding bending moment directions, the resultant net bending moment, and the scrolling direction for MXene with different surface terminations in Class II:CPSF: (a) $bTi_2C(O)$, (b) $bTi_2C(OH)$, and (c) $(O)Ti_2C(OH)$.**

### 3.5. Single Ti nanoparticle loaded onto plane strain MXene (Class IV:TiNP@bTi₂C(OH))

One of the most promising strategies for developing high-capacity anodes is to combine an active material with a coating system to form a core@shell composite structure [47–50]. In this type of composite, the active material serves as the core. For example, silicon (Si) and sulfur (S) are widely used as core materials for lithium-ion (Li-ion) and sodium-ion (Na-ion) batteries [51–53], respectively, because they provide high charge-storage capacities through alloying reactions. However, these materials undergo significant volume expansion (approximately 300% for Si [54] and 80% for S [55]) during charge/discharge cycles, leading to anode pulverization and reduced cycle life. To overcome this limitation, a coating material with high mechanical durability is used to form the shell of the core@shell composite [47–50]. The shell accommodates volume expansion, maintains the electrode's structural integrity, and provides electronic conductivity.

Core@shell composites combine the high storage capacity of the core material with the controlled volume expansion and enhanced electronic conductivity provided by the shell, making them promising electrode architectures for next-generation energy storage systems. However, one of the primary challenges associated with these composites is the shell-coating strategy, which can restrict the electrochemical capacity of the active core material [56,57]. In practice, the core material requires sufficient space to accommodate the volume changes associated with lithiation/delithiation or sodiation/desodiation during charge/discharge cycles. Consequently, if the shell excessively constrains the core, the accessible capacity of the active material is reduced. To overcome this limitation, a scrolled shell configuration may provide an effective alternative. Unlike a rigid shell, a scrolled shell can accommodate repeated expansion and contraction in a spring-like manner during lithiation/sodiation and delithiation/desodiation, respectively. As a result, the active material can undergo reversible volume changes without being excessively constrained, thereby preserving its electrochemical capacity while maintaining the mechanical integrity of the composite.

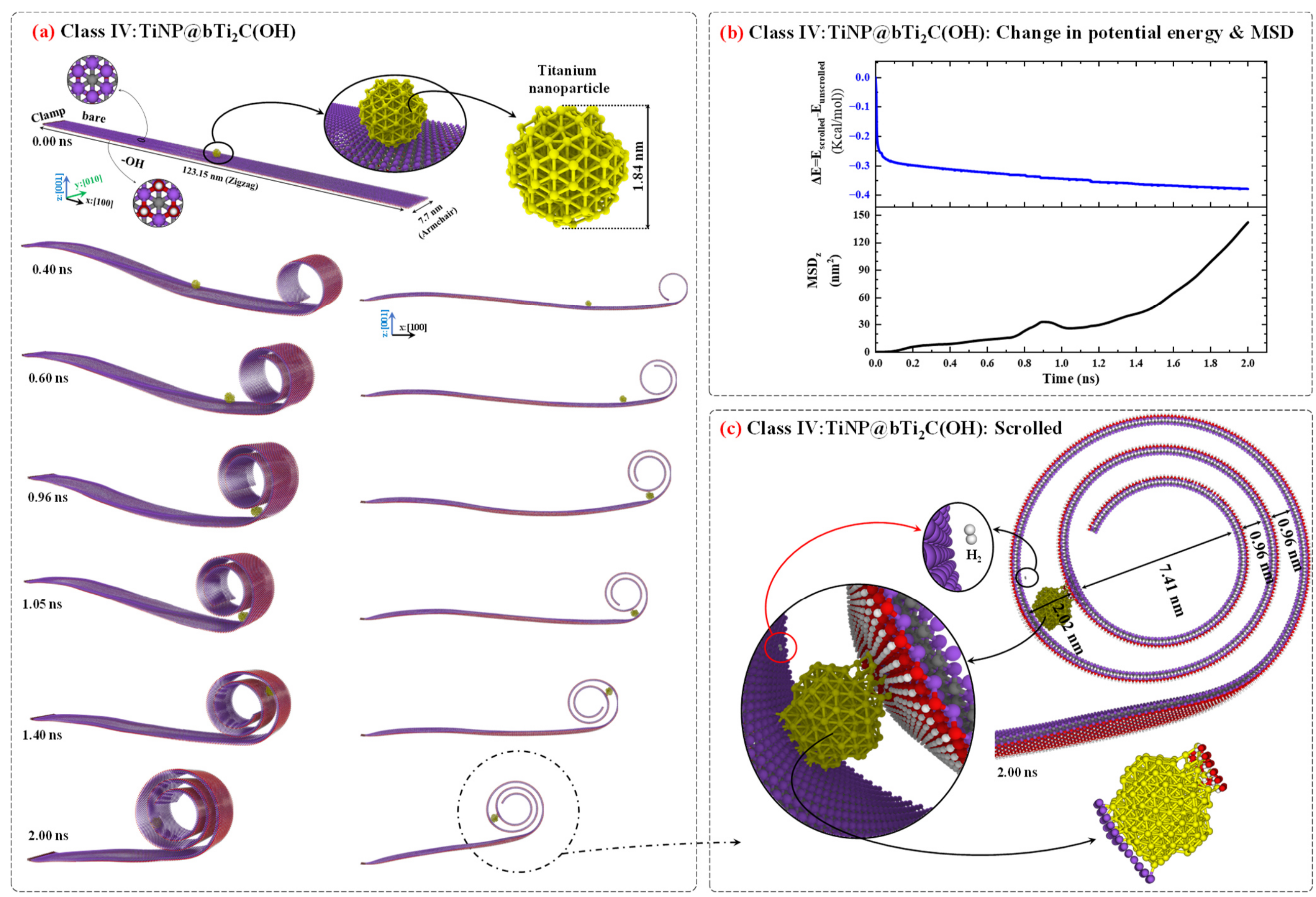

**Figure 8 Self-scrolling behavior, energy, and MSD diagrams of the Class IV:TiNP@bTi₂C(OH). (a) Scrolling snapshots of Class IV: CPSF: TiNP@bTi2C(OH). (b) Evolution of the potential energy. (c) Mean squared displacement (MSD) along the z:[001] direction during the 1 ns simulation.**

In light of these challenges, nanoscrolled MXene appears to be a promising candidate for use as a rolled shell to protect active nano-materials. However, this application requires that the MXene retain its self-scrolling capability in the presence of active nano-materials. To evaluate this capability, a TiNP was introduced as a model spherical active material and placed on the bare surface of Class IV:b$Ti_2C(OH)$ at the center of the sheet. The self-scrolling behavior of the TiNP@b$Ti_2C(OH)$ system was then investigated, as illustrated in Figure 8. As shown in this figure, the overall results demonstrate that the self-scrolling mechanism is preserved in the presence of the TiNP. Figure 8.a clearly shows that the scrolling behavior of TiNP@b$Ti_2C(OH)$ is similar to that of pristine b$Ti_2C(OH)$, with the sheet rolling toward the -b surface termination. Furthermore, the potential energy per atom and MSD diagrams as a function of time (Figure 8.b) exhibit trends similar to those observed for pristine b$Ti_2C(OH)$. However, at approximately 0.96 ns, the MSD curve exhibits an abrupt peak, indicating that the scrolled region of the MXene sheet comes into contact with the TiNP. The nanoparticle is subsequently captured between the concentric MXene layers, after which the scrolling process continues until the end.

Figure 8.c presents a magnified view of the nanoscrolled TiNP@b$Ti_2C(OH)$ system at 2 ns. The figure clearly demonstrates that the presence of the TiNP alters the otherwise uniform interlayer spacing of the scrolled MXene sheets. In addition, the enlarged view of the TiNP confined between two adjacent MXene layers reveals that the Ti atoms of the nanoparticle ($Ti_{TiNP}$) form chemical bonds with the Ti and O atoms of the MXene ($Ti_{MXene}$ and $O_{MXene}$), thereby stabilizing the TiNP within the scrolled structure. Furthermore, the chemical interaction between the $Ti_{TiNP}$ and the $O_{MXene}$ weakens the O-H bonds on the -OH surface termination, resulting in the release of H atoms, which subsequently recombine to form hydrogen molecules ($H_2$), as illustrated in Figure 8.c. This H evolution originates from the stronger affinity of $O_{MXene}$ (3.5) for the electrons of $Ti_{TiNP}$ compared with H (2.2) [58], owing to the higher electronegativity of $O_{MXene}$. Consequently, $O_{MXene}$ preferentially forms Ti-O bonds with the TiNP, reducing its interaction with H and promoting the dissociation of O-H bonds. The released H atoms subsequently combine to form $H_2$ molecules. If these $H_2$ remain confined within the nanoscrolled MXene layers and are introduced into an electrochemical cell, they may accumulate as $H_2$ bubbles at the electrode surface. Such bubbles can block electrochemically active sites, hinder ion transport, increase ohmic resistance, and degrade the battery's electrochemical performance. This phenomenon, commonly referred to as the “$H_2$ bubble shielding effect”, is generally detrimental to battery performance [59–61].

Table 6 compares the interlayer spacing of the nanoscrolled MXene sheets at locations far from the TiNP and in the vicinity of it. The results show that the interlayer spacing increases from 0.96 nm in regions distant from the TiNP to 2.02 nm at the TiNP location. The larger interlayer spacing at the TiNP site exceeds the initial diameter of the TiNP (1.84 nm) owing to van der Waals interactions and the formation of chemical bonds between the $Ti_{TiNP}$ and the $Ti_{MXene}$ and $O_{MXene}$. Furthermore, the diameter of the innermost scroll is 7.41 nm, which is close to the value obtained for pristine b$Ti_2C(OH)$, as reported in Table 4. These results demonstrate that the interlayer spacing of nanoscrolled MXene can be effectively tuned by incorporating nanoparticles. This tunability represents a suitable advantage because varying the nanoparticle size enables precise control of the spacing between adjacent MXene layers. Such controllable interlayer spacing is expected to facilitate $Li^+$/$Na^+$ intercalation, enhance the charge-storage capacity of the MXene anode, and ultimately improve the overall electrochemical performance and capacity of the battery.

**Table 6 Comparison of the interlayer distance and the diameter of the innermost ring of MXene nanoscroll in Class IV:CPSF: TiNP@bTi$_2$C(OH). The initial diameter of the TiNP was 1.84 nm.**

| Class | Surface termination | Interlayer distance (nm) measured far from the Ti nanoparticle | Interlayer distance (nm) measured close to the Ti nanoparticle | Diameter of innermost ring (nm) | Stability status |
|---|---|---|---|---|---|
| IV | bTi$_2$C(OH) | 0.96 | 2.02 | 7.41 | Stable |

### 3.6. Limitations and Perspective

Within the scope of this study, because the available ReaxFF force field [38] does not include reactive parameters for Si or S, the active nanoparticle was modeled using a TiNP. This approach enabled the evaluation of the nanoscrolling capability in the presence of a nanoparticle, with the TiNP serving as a model representative of Si and S nanoparticles. However, it should be acknowledged that the interaction of Ti with MXene differs from that of Si and S. For practical applications, the active nanoparticles should be composed of commercially relevant electrode materials, such as Si or S, and the possibility of $H_2$ bubble formation should be systematically investigated. In this regard, future studies should focus on realistic Si- or S-based nanoparticles by developing appropriate ReaxFF force fields and examining the formation of $H_2$ nanobubbles in Si (or S)@MXene/electrolyte systems using reactive molecular dynamics simulations.

Because of the scrolled morphology of MXene, if this material is used as the shell in a core@shell composite anode for energy storage devices, spring-like expansion/contraction can be predicted during charge/discharge cycles. However, this prediction still requires validation. Reactive MD simulations of Si (or S)@MXene/electrolyte systems, or Si (or S)@MXene systems in contact with a Li source (or a Na source), can be employed for this purpose. In these systems, applying an external voltage induces lithiation/delithiation or sodiation/desodiation, enabling evaluation of the proposed spring-like expansion/contraction behavior. Furthermore, the effects of nanoparticle configuration (e.g., nanoparticles, nanorods, and nanowires) and particle size on the nanoscrolling ability of MXene represent additional promising research topics that should be investigated to understand better the influence of nanomaterial morphology and dimensions on the scrolling mechanism.

For Class III: FSS, we investigated the square configurations with a width-to-length ratio of 1 and sizes up to 100 nm. This ratio and the initial flake size affect the nanoscrolling behavior and the final scrolled configuration [35]. In the laboratory, the MXene flakes are usually of micron scale. To understand the effects of the width-to-length ratio and initial size, future studies should investigate different MXene structures with different width-to-length ratios and sizes. Such studies can reveal the morphology of nanoscrolled MXene as a function of size, width, length, and scrolling direction. The results obtained from these investigations can provide valuable benchmarks for selecting appropriate MXene nanoscrolls for battery materials.

Beyond the direct attributes related to energy storage applications, such as morphology and expansion/contraction behavior during charge/discharge cycles, the indirect properties of nanoscrolled MXene, including mechanical characteristics, thermal expansion and stability, and vibrational properties, also require careful consideration [62,63]. Regarding mechanical properties, nanoscrolled MXene should provide sufficient mechanical stability for use as an anode material to prevent fracture during cell assembly and deformation under external loading. In terms of thermal properties, nanoscrolled MXene should remain thermally stable within the anode's operating temperature range without significant thermal expansion, thereby maintaining effective surface contact with the binder. Concerning vibrational properties, nanoscrolled MXene should exhibit low-frequency vibrational modes because thermal expansion is dependent on the frequency of normal modes [63,64]. In this regard, high-frequency modes can lead to significant thermal expansion and mechanical degradation. Therefore, the mechanical, thermal, and vibrational properties of nanoscrolled MXene should be systematically investigated to ensure the required performance and establish appropriate criteria for selecting MXene structures as anode materials for energy storage devices.

## 4. Conclusions

This study investigates the nanoscrolling mechanisms, morphologies, and potential applications of MXene under four simulation classes: Class I: FPSF, Class II: CPSF, Class III: FSS, and Class IV:CPSF:TiNP:b$Ti_2C(OH)$, with different surface terminations of b$Ti_2C(O)$, b$Ti_2C(OH)$, and (O)$Ti_2C(OH)$. The primary objective of this study is to provide insights into the mechanisms of nanoscroll formation and particle encapsulation in Janus MXenes. To this end, the initial configurations of all MXene systems were constructed based on experimental reports, and the nanoscrolling phenomena were systematically investigated using MD simulations.

The results reveal that surface etching of MXene, which produces the -b, -O, and -OH surface terminations, alters the bond lengths at the surface. Consequently, tensile and compressive strains are generated on the MXene surfaces depending on the corresponding surface functional groups. This strain mismatch induces a net bending moment that drives the nanosheet to scroll toward the surface, where it experiences compressive strain. Accordingly, in all simulation classes, the MXene systems with b$Ti_2C(O)$, b$Ti_2C(OH)$, and (O)$Ti_2C(OH)$ surface terminations bend toward the -O, -b, and -O surfaces, respectively, because these surfaces experience more compressive strains relative to the strains developed on the opposite surfaces.

The investigation of MXene with two free ends (Class I:FPSF) and one clamped end (Class II:CPSF) demonstrates that both the surface termination and the clamping condition influence the nanoscrolling mechanism and the resulting morphology. Among the surface terminations considered in this study, b$Ti_2C(O)$ forms a metastable semi-nanoscrolled structure and, due to its limited structural stability, is not suitable for practical applications such as anode materials for energy storage systems. In contrast, MXene with b$Ti_2C(OH)$ and (O)$Ti_2C(OH)$ surface terminations forms stable nanoscrolls and are recommended for practical applications. Clamping one end of the MXene sheet slightly increases both the scrolling diameter and the interlayer distance. Furthermore, the (O)$Ti_2C(OH)$ system produces larger nanoscrolls than the b$Ti_2C(OH)$ system. The average diameter of the innermost ring and the interlayer distance are 6.25 nm and 0.80 nm, respectively, for b$Ti_2C(OH)$, whereas the corresponding values for (O)$Ti_2C(OH)$ are 7.50 nm and 0.74 nm.

Investigating Class III:FSS MXene demonstrates that three distinct morphologies of nanocurved, nanotube, and nanoscrolled MXene can be achieved by manipulating the surface termination and sheet size. In this class, the asymmetric surface termination generates the bending moment, which acts as the driving force for bending, while the sheet size controls the final morphology of the curved MXene. For FSS MXene sheets with sizes smaller than 20 nm, only the nanocurved morphology can be observed due to size limitations. However, $bTi_2C(OH)$ is an exception because the curvature induced by the bending moment reaches a suitable diameter that enables the formation of a chiral MXene nanotube. For sheet sizes larger than 20 nm, the possibility of nanoscrolling increases, and nanoscrolled $bTi_2C(OH)$ and $(O)Ti_2C(OH)$ structures can be obtained under FSS configurations. In all the studied square flakes with up to 100 nm size, $bTi_2C(O)$ can only form a nanocurved configuration.

To evaluate the feasibility of fabricating core@shell composites with a nanomaterial-based core and a nanoscrolled MXene shell, the self-scrolling behavior of $bTi_2C(OH)$ was investigated in the presence of a TiNP. The results demonstrate that MXene retains its self-scrolling capability in the presence of the TiNP and successfully encapsulates the nanoparticle between its concentric MXene layers. Furthermore, the interlayer spacing of the nanoscrolled MXene can be tuned by varying the nanoparticle size. However, the simulations also reveal the possibility of $H_2$ release and subsequent $H_2$ nanobubble formation. The ability to achieve self-scrolling and controllable interlayer spacing through nanoparticle size represents an advantage for electrode design. The nanoscrolled shell can accommodate expansion/contraction in a spring-like manner during charge/discharge cycles, while its controllable interlayer spacing can enhance the $Li^+$ or $Na^+$ intercalation capacity. However, $H_2$ release and the potential formation of $H_2$ nanobubbles may adversely affect the electrochemical performance of the anode because $H_2$ nanobubbles can lead to the bubble shielding effect, thereby blocking electrochemically active sites, hindering ion transport, and increasing the ohmic resistance of the battery.

## Data availability

Data are available from the corresponding authors upon reasonable request.

## Author Contributions

**Sasan Rezaee**: Conceptualization, Methodology, Data curation, Formal analysis, Investigation, Software, Validation, Visualization, Writing-original draft, Writing-review, and Editing. **Fatemeh Mohammad Dezashibib**: Investigation, Writing-original draft, Writing-review, and Editing. **Ould el Moctar**: Conceptualization, Supervision, Writing-original draft, Writing-review, and Editing. **Hossein Darband**: Conceptualization, Project administration, Supervision, Validation, Methodology, Data curation, Investigation, Software, Formal analysis, Writing-original draft, Writing-review, and Editing.

## Declaration of Interests

The author declares that he has no known competing financial interests or personal relationships that could have appeared to influence the work reported in this paper.

## Funding

This research did not receive any specific grant from funding agencies in the public, commercial, or not-for-profit sectors.

## Acknowledgment

The authors gratefully acknowledge Polish high-performance computing infrastructure PLGrid (HPC Centers: WCSS, ACK Cyfronet AGH) for providing computer facilities and support within computational grant no. PLG/2026/019291.